\documentclass[aps,prl,twocolumn,showpacs,preprintnumbers,amsmath,amssymb]{revtex4-1}
\usepackage{latexsym}
\usepackage{graphicx}
\usepackage{graphics}
\usepackage[T1]{fontenc}
\usepackage{multirow}
\usepackage[british,UKenglish]{babel}
\usepackage{color}
\usepackage{ulem}

\newcommand{\stkout}[1]{\ifmmode\text{\sout{\ensuremath{#1}}}\else\sout{#1}\fi}

\begin{document}

\author{Florian Kaiser$^{1,\dagger}$}
\email{florian.kaiser@unice.fr}
\author{Panagiotis Vergyris$^1$, Djeylan Aktas$^1$,\\Charles Babin$^{1,2}$, Laurent Labont\'e$^1$}
\author{S\'ebastien Tanzilli$^1$}
\affiliation{$^1$Universit\'e C\^{o}te d'Azur, Institut de Physique de Nice (INPHYNI), CNRS UMR 7010, Parc Valrose, 06108 Nice Cedex 2, France\\
$^2$\'Ecole Normale Sup\'erieure de Lyon, 46 All\'ee d'Italie, 69364 Lyon Cedex 07, France\\
$^{\dagger}$ Now at 3. Physikalisches Institut, Universit\"at Stuttgart, Pfaffenwaldring 57, 70569 Stuttgart, Germany}

\title{Quantum enhancement of accuracy and precision in optical interferometry}

\begin{abstract}
White-light interferometry is one of today's most precise tools for determining optical material properties. Achievable precision and accuracy are typically limited by systematic errors due to a high number of interdependent data fitting parameters.
Here, we introduce spectrally-resolved quantum white-light interferometry as a novel tool for optical property measurements, notably chromatic dispersion in optical fibres. By exploiting both spectral and photon-number correlations of energy-time entangled photon pairs, the number of fitting parameters is significantly reduced which eliminates systematic errors and leads to an absolute determination of the material parameter.
By comparing the quantum method to state-of-the-art approaches, we demonstrate the quantum advantage through 2.4 times better measurement precision, despite involving 62 times less photons.
The improved results are due to conceptual advantages enabled by quantum optics which are likely to define new standards in experimental methods for characterising optical materials.
\end{abstract}

\maketitle

\section{Introduction}

Quantum technologies received great attention as means to improve resolution and precision of metrological tasks by reducing statistical errors due to quantum noise~\cite{Giovannetti_QuMetReview_2004,Walther_4photon_2004,Mitchell_superresolution_2004,Nagata_superresolution_2007,Higgins_entanglement-free_2007,Matthews_multiphoton-wg_2009,Afek_high-noon_2010,Advances_in_Qmet}. Far less attention has been paid to their ability to reduce systematic errors. However, statistical and systematic errors are of equal importance in any measurement, and the latter ones are typically more difficult to account for.
Notable examples for quantum-improved measurements are the use of single-electron sources for a more accurate definition of the Ampere~\cite{Poirier_ampere_2016}, and quantum correlated ``twin photon beams'' towards establishing absolute and universal optical power standards~\cite{Migdall_power_1995}.
In this letter we demonstrate a new use of quantum optics to reduce systematic errors in the technologically prominent application of spectrally-resolved white-light interferometry (WLI). WLI is used for precise measurements of chromatic dispersion, \textit{i.e.} the second derivative of the wavelength-dependent optical phase.
Classical WLI requires, however, precise interferometer equalisation~\cite{Naganuma_subwavelength_1990,Diddams_dispersion-WLI_1996} and is influenced by third-order dispersion~\cite{MethodsComparison,Galle_thesis_2014} which leads to systematic errors that are difficult to account for.\\
We eliminate all those drawbacks by inferring chromatic dispersion using energy-time entangled photon pairs and coincidence counting to measure spectral correlation functions. In addition, we exploit photon-number correlations to achieve a two-fold resolution enhancement.
Our results demonstrate that this new strategy outperforms the precision and accuracy of previous quantum~\cite{Brendel_dispersion_1998,Nasr_dispersion_2004} and state-of-the-art techniques~\cite{Naganuma_subwavelength_1990,Diddams_dispersion-WLI_1996}. Moreover, as our approach is essentially alignment-free, it enables using the same interferometer in a user-friendly fashion for analysing a wide variety of different optical materials, in terms of type, optical properties, length, \textit{etc.}.

\subsection{Standard white-light interferometry}

\begin{figure}
\includegraphics{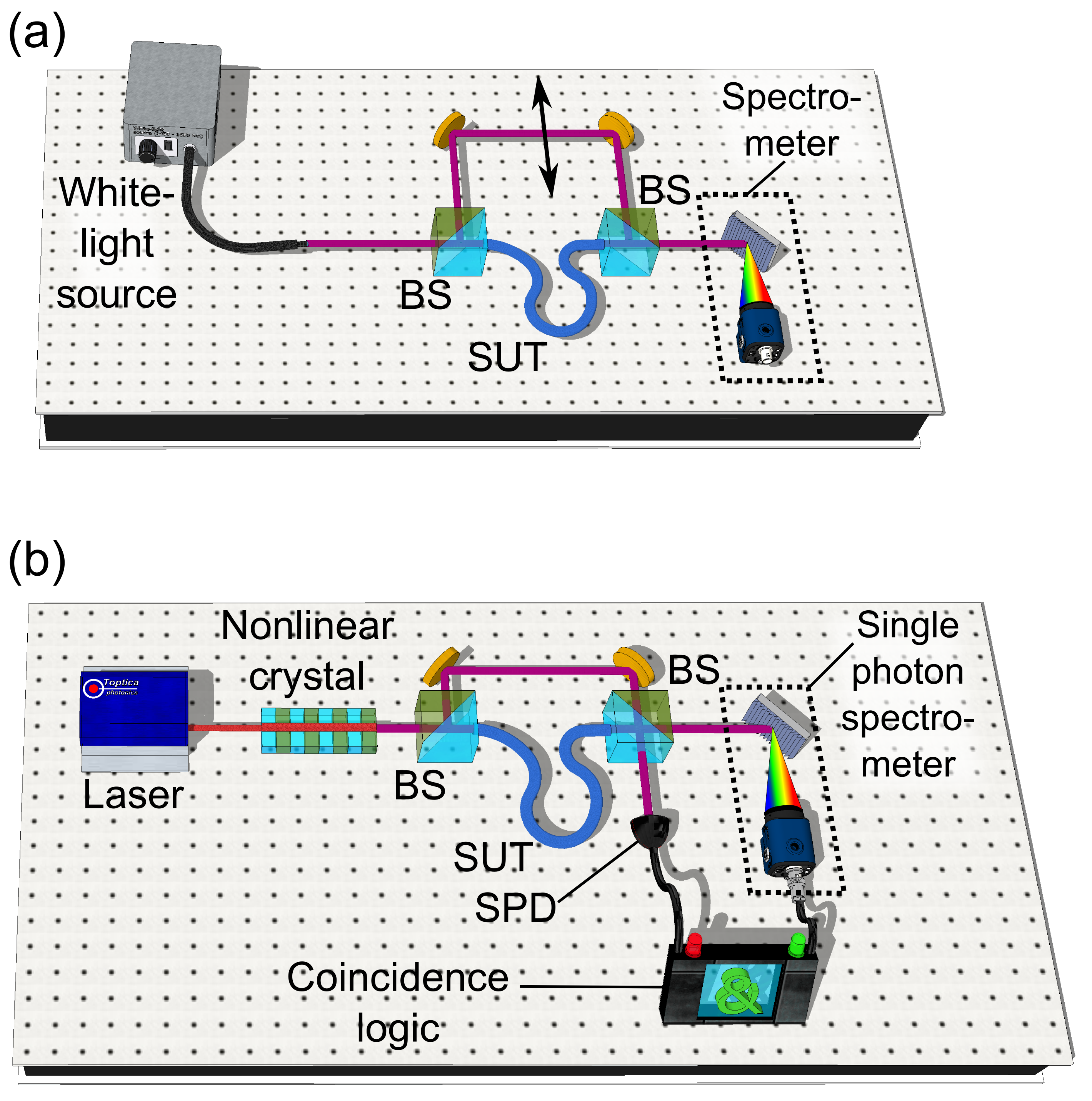}
\caption{Typical experimental setups for standard spectrally-resolved WLI \textbf{(a)}, and Q-WLI \textbf{(b)}. BS, beam-splitter; SUT, sample under test; SPD, single photon detector. \&-symbol, time-tagging and coincidence logic.\label{Fig1}}
\end{figure}

The standard scheme for WLI is shown in \figurename~\ref{Fig1}(a). The emission of a white-light source is directed to an interferometer in which the reference arm is free-space (with well known optical properties) and the other arm comprises the sample under test (SUT).
Recombining both arms at the output beam-splitter leads to an interference pattern for which the intensity follows $I \propto 1 + \cos \left( \phi(\lambda) \right)$, with
$\phi(\lambda) = \tfrac{2\,\pi}{\lambda} \left(n(\lambda) \cdot L_{\rm s} - L_{\rm r} \right)$. Here, $\lambda$ represents the wavelength, $L_{\rm r}$ and $L_{\rm s}$ are the physical lengths of the reference arm and the SUT, respectively, and $n(\lambda)$ is the effective refractive index of the SUT.
It is worth noting that interference is only observed when the interferometer is precisely balanced to within the larger of: the coherence length of the white-light source and the coherence length imposed by the resolution of the spectrometer, which is typically on the order of microns to millimetres~\cite{Naganuma_subwavelength_1990,Diddams_dispersion-WLI_1996}.
In this case, the phase term reads (more details are given in the supplementary information):
\begin{eqnarray}
\phi(\lambda) &\approx& 2\,\pi\,L_{\rm s} \Bigg( \frac{1}{2} \, \frac{d^2 n}{d \lambda^2}\Bigg|_{\lambda_0} \cdot \frac{(\Delta \lambda) ^2}{\lambda_0  + \Delta \lambda} \nonumber\\
&\,& + \frac{1}{6} \, \frac{d^3 n}{d \lambda^3}\Bigg|_{ \lambda_0} \cdot \frac{(\Delta \lambda) ^3}{\lambda_0 + \Delta \lambda} \Bigg) + \phi_{\rm off}. \label{ClassicalPhase}
\end{eqnarray}
Here, $\lambda_0$ represents the stationary phase point, \textit{i.e.} the wavelength at which the absolute phase difference between the two interferometer arms is exactly zero. In standard WLI, $\lambda_0$ is extracted experimentally by identifying the symmetry point of the observed interferogram~\cite{Naganuma_subwavelength_1990,Diddams_dispersion-WLI_1996}. Additionally, $\Delta \lambda = \lambda - \lambda_0$, and $\phi_{\rm off}$ is a constant offset phase.
Provided that $L_{\rm s}$ is precisely known, the optical material parameters $\frac{d^2n}{d \lambda^2}\Big|_{\lambda_0}$ and $\frac{d^3n}{d \lambda^3}\Big|_{\lambda_0}$ can be extracted from a fit to the data as a function of $\Delta \lambda$. It is noteworthy that the three free parameters, \textit{i.e.} $\lambda_0$, $\frac{d^2n}{d \lambda^2}\Big|_{\lambda_0}$ and $\frac{d^3n}{d \lambda^3}\Big|_{\lambda_0}$, are usually all interdependent in a non-trivial fashion such that uncertainties on one parameter systematically induce uncertainties on the others.
As a matter of fact, the high number of required fitting parameters and the necessity to re-equilibrate the interferometer for every new SUT represent the main limiting factors of this technique~\cite{MethodsComparison,Galle_thesis_2014}.

However, more accurate optical measurements are eagerly demanded in almost all fields where optics is involved. A special focus is set on the optical parameter $\tfrac{d^2n}{d \lambda^2}\Big|_{\lambda_0}$ as it is directly related to the chromatic dispersion coefficient $D = -  \frac{\lambda_0}{c} \cdot \frac{d^2n}{d \lambda^2}\Big|_{\lambda_0}$, in which $c$ represents the speed of light~\cite{Hlubina_one_percent,Laurent_dispersion,Kardas_160fs,Hlubina_sub_100fs,Ye_few_percent,MethodsComparison,Galle_280fs,GDD_mirrors}.
Chromatic dispersion causes optical pulse broadening and more accurate measurements on $D$ would have tremendous repercussions for optimising today's telecommunication networks, developing new-generation pulsed lasers and amplifiers, designing novel linear and nonlinear optical components and circuits, as well as for assessing the properties of biological tissues.

\section{Materials and Methods}

\subsection{Quantum white-light interferometry}

\figurename~\ref{Fig1}(b) depicts the new experimental schematic dedicated to spectrally-resolved quantum WLI (Q-WLI), intended to overcome the above-mentioned issues.
The quantum white-light source is composed of a continuous-wave pump laser and a non-linear crystal in which energy-time entangled photon pairs are generated through spontaneous parametric downconversion~\cite{Kaiser_source,Alibart_PPLN_2016}.
This process obeys the conservation of the energy, \textit{i.e.} $\frac{1}{\lambda_{\rm p}} = \frac{1}{\lambda_1} + \frac{1}{\lambda_2}$. Here, $\lambda_{\rm p,1,2}$ represent the vacuum wavelengths of the pump laser photons, and the individual photons for each generated pair. Another implication of the conservation of the energy is that the degenerate vacuum wavelength of the emission spectrum is $\lambda^* = 2\,\lambda_{\rm p}$.
We send the paired photons to the interferometer, however, as opposed to standard WLI, we intentionally unbalance it now.
This provides us with two advantages: first, we avoid single-photon interference; and second, we obtain a means to distinguish events where both photons take opposite paths (strongly delayed arrival times at the interferometer's outputs) or the same path (near zero arrival time difference)~\cite{Kaiser_source}. We post-select the latter ones by considering only two-photon coincidence detection events in which both the single photon detector (SPD) and the single-photon sensitive spectrometer fire simultaneously.
Our target is now to observe quantum interference between those two-photon contributions which necessitates that they are coherent and indistinguishable. Coherence is ensured by operating the interferometer at a path length difference that is shorter than the coherence length of the pump laser ($\sim$100\,m) such that the photon-pair contributions \textit{are in phase}~\cite{Franson_energy_time_1989}.
Indistinguishability concerns mainly the temporal envelope of the photon-pair wave packet which is distorted from its original shape by the dispersion-induced temporal walk-off between the individual photons in the SUT. For standard fibres, this means that path length differences up to $\sim$10\,m are acceptable~\cite{Vergyris_sagnac_2017}.\\
Thus, provided that the interferometer is operated in these conditions, near zero arrival time coincidence detection results in a two-photon $N00N$-state:
\begin{equation}
|\psi \rangle = \frac{\left( |2 \rangle_{\rm r} |0 \rangle_{\rm s} + {\rm e}^{{\rm i} \phi_{N00N}} |0 \rangle_{\rm r} |2 \rangle_{\rm s} \right)}{\sqrt{2}}.
\end{equation}
Here, the ket vectors, indexed by s and r, indicate the number of photons in the reference and SUT arm, respectively, and $\phi_{N00N} = \phi(\lambda_1) + \phi(\lambda_2)$.
We obtain the spectral dependence of $\phi_{N00N}$ by computing $\phi(\lambda_1)$ and $\phi(\lambda_2)$ accordingly to equation~\ref{ClassicalPhase} and respecting the conservation of the energy during the downconversion process:
\begin{equation}
\phi_{N00N}(\lambda) \approx \frac{d^2 n}{d \lambda^2}\Bigg|_{\lambda^*} \cdot \frac{\pi \, L_{\rm s} \cdot  (\Delta \lambda)^2}{\frac{\lambda^*}{2}+\Delta \lambda} + \phi_{\rm off}. \label{QuantumPhase}
\end{equation}
Here, $\phi_{\rm off} = \frac{4\,\pi \left( n(\lambda^*) \,L_{\rm s} - L_{\rm r} \right) }{\lambda^*}$ is an offset term, and $\Delta \lambda = \lambda - \lambda^*$. The phase-dependent two-photon coincidence rate $R$ is then: $R \propto 1+\cos \left( \phi_{N00N} \right)$.
In the past, numerous studies have investigated the term $\phi_{\rm off}$, as it allows measuring optical phase-shifts at constant wavelengths with doubled resolution compared to the standard approach~\cite{Crepsi_protein-concentration_2012,Ono_microscopy_2013,Israel_microscopy_2014}.\\
We access here, for the first time, the wavelength-dependent term in equation~\ref{QuantumPhase} by recording $R$ as a function of $\Delta \lambda$, \textit{i.e.} the two-photon coincidence rate is measured as a function of the paired-photons' wavelengths.

This leads to several pertinent purely quantum-enabled features:
due to the use of an energy-time entangled two-photon $N00N$-state, the required precision on equilibrating the interferometer is $\sim$10\,m instead of microns to millimetres in standard WLI~\cite{Naganuma_subwavelength_1990,Diddams_dispersion-WLI_1996,MethodsComparison,Galle_thesis_2014}. This is particularly interesting for improving the ease-of-use as no re-alignment is necessary when changing the SUT;
compared to equation~\ref{ClassicalPhase}, the third-order term $\tfrac{d^3n}{d \lambda^3}$ in equation~\ref{QuantumPhase} is cancelled thanks to energy-time correlations~\cite{Nasr_QOCT_2004};
furthermore, the wavelength at which chromatic dispersion is measured, $\lambda^*$, does not have to be extracted from the data, as it is exactly twice the wavelength of the continuous-wave pump laser, $\lambda_{\rm p}$, and can therefore be known with extremely high accuracy.
This means that the quantum strategy allows data fitting using exactly one free parameter, namely $\tfrac{d^2n}{d \lambda^2}\Big|_{\lambda^*}$ which is an essential step towards absolute optical property determination with high precision without systematic errors.
Finally, due to the use of a two-photon $N00N$-state, a doubled resolution on $\tfrac{d^2n}{d \lambda^2}\Big|_{\lambda^*}$ is achieved, allowing to perform measurements on shorter samples and components compared to standard WLI, \textit{i.e.} down to the technologically interesting mm to cm scale.

\subsection{Detailed optical setup and data acquisition}

To benchmark standard and quantum approaches, we employ a 1\,m long SMF28e fibre from Corning as SUT.
For all measurements, we employ the same interferometer and actively stabilise it using a reference laser and a piezoelectric transducer on one mirror in the reference arm (more details are provided in the methods section). This ensures that $\phi_{\rm off}$ remains constant.\\
For chromatic dispersion measurements using classical WLI, we use a state-of-the-art superluminescent diode. At the output of the interferometer we measure an average spectral intensity of $\sim$125\,pW/nm from $1450$ to $1650\rm\,nm$. Interferograms are recorded using a spectrometer from Anritsu (model MS9710B) with 0.1\,s integration time and 0.5\,nm resolution which are standard parameters for this kind of measurement~\cite{Naganuma_subwavelength_1990,Diddams_dispersion-WLI_1996}.\\
For the Q-WLI approach, the light source is made of a 780.246\,nm laser, pumping a \mbox{type-0} periodically-poled lithium niobate waveguide (PPLN/W). We stabilise the laser wavelength against the $F=2 \rightarrow F'=2 \times 3$ hyperfine crossover transition in atomic $^{87}$Rb, such that $\lambda_{\rm p}$ and $\lambda^*$ are known with a precision on the order of 1\,fm.
The quasi-phase matching in the PPLN/W is engineered such as to generate energy-time entangled photon pairs around the degenerate wavelength of $\lambda^*=1560.493\rm\,nm$ with a bandwidth of about 140\,nm~\cite{Alibart_PPLN_2016}.
To detect the paired photons, we use an InGaAs SPD (IDQ 220) at one interferometer output. The single photon spectrometer at the other output is made of a wavelength-tunable 0.5\,nm bandpass filter, followed by another InGaAs SPD (IDQ 230).
To avoid saturation of these detectors, the spectral intensity at the interferometer output is reduced to about 25\,fW/nm, which is partially compensated by increasing the integration time to 8\,s.\\
All measurements are repeated 100 times on the same SUT in order to infer the statistical accuracy of both WLI and Q-WLI approaches.

\section{Results and discussion}

\subsection{Statistical analysis for comparing measurement precision}

\begin{figure}[h] 
\includegraphics{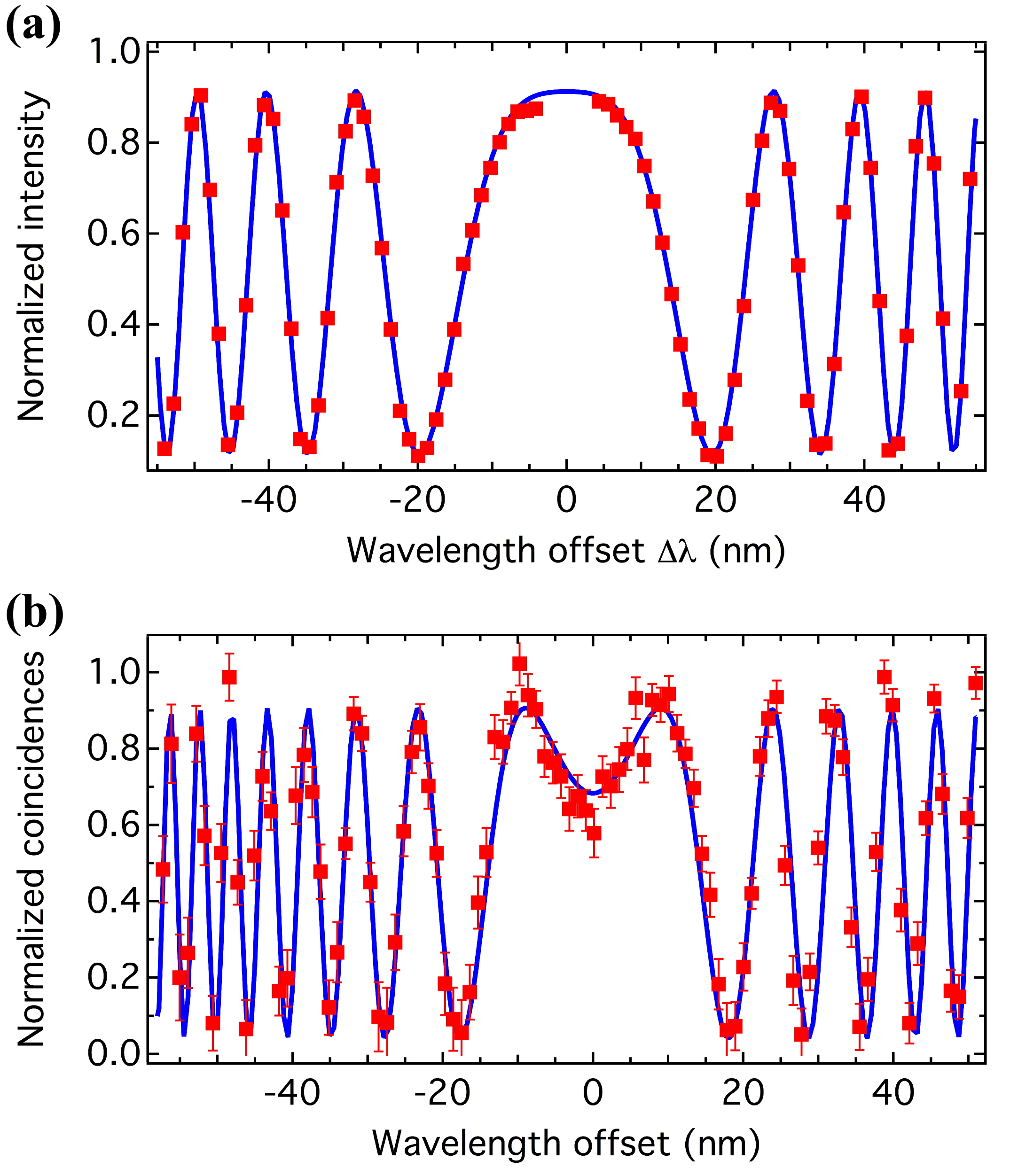}
\caption{Typical measurements acquired for inferring chromatic dispersion in a 1\,m long standard single-mode fibre using standard WLI \textbf{(a)}, and Q-WLI \textbf{(b)}.
Red dots, data points; Blue lines, appropriate fits to the data following equations~\ref{ClassicalPhase} and \ref{QuantumPhase} from which $D$ is extracted. Error bars assume poissonian photon number statistics. For standard WLI, normalization is obtained by measuring two reference spectra. For Q-WLI, normalization is performed \textit{on-the-fly} by counting non-zero arrival time difference coincidences. For more details, refer to the supplementary information.\label{Fig2}}
\end{figure}

\begin{center}
\begin{figure}[h] 
\includegraphics{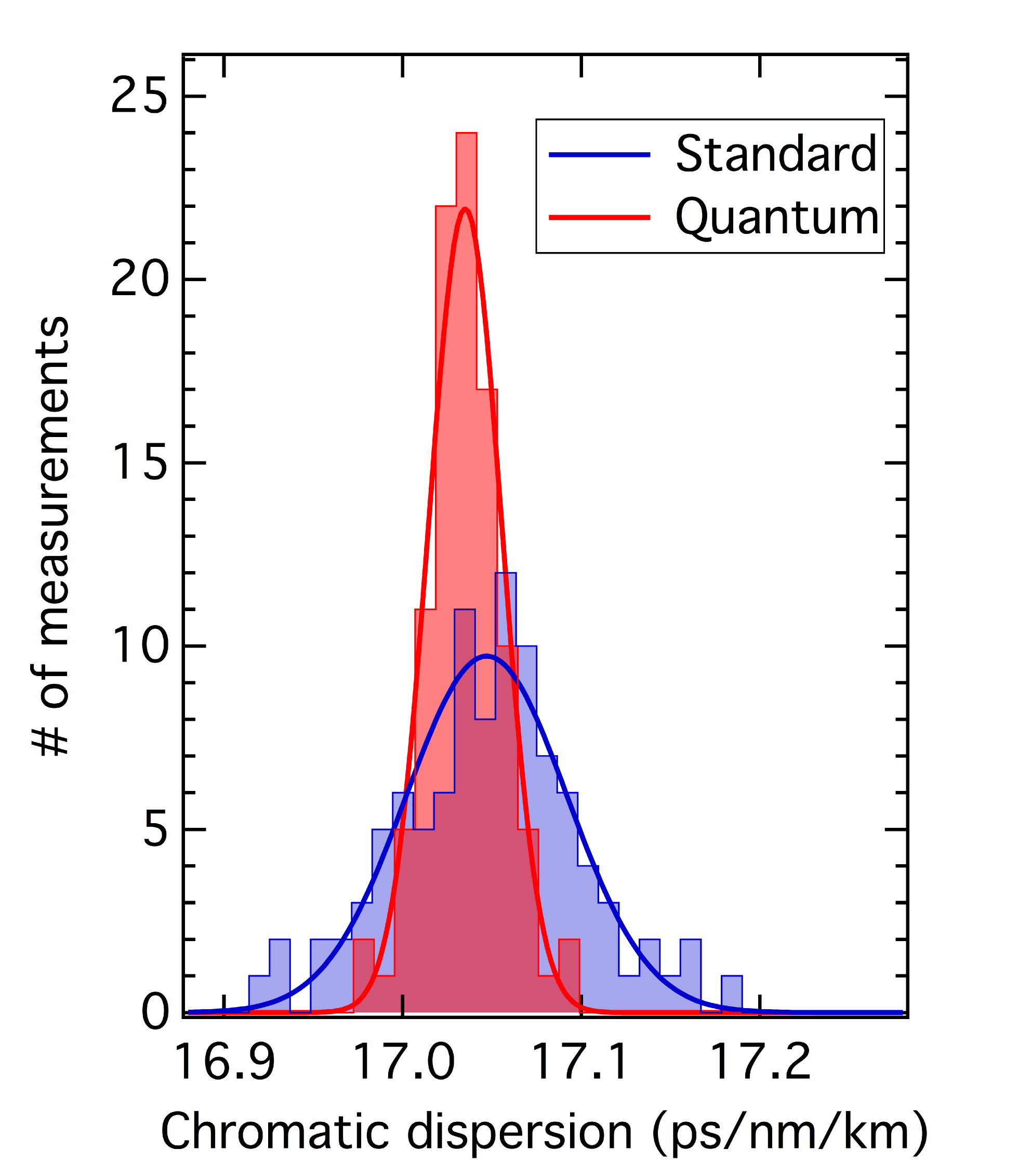}
\caption{Histogram of inferred chromatic dispersion coefficients after 100 repetitions on the same SUT for both standard (blue) and quantum-enhanced (red) measurements, respectively. Fits to the data assume a normal distribution.\label{Fig3}}
\end{figure}    
\end{center}

Typical interference patterns for chromatic dispersion measurements using both methods are shown in \figurename~\ref{Fig2}(a,b).
With the Q-WLI setup, we find twice as much interference fringes for the same spectral bandwidth which is a direct consequence of the doubled phase sensitivity of the two-photon $N00N$-state.
After acquiring $2 \times 100$ measurements on the same SUT, we infer the precision of both approaches. The results of the statistical data analysis are shown in \figurename~\ref{Fig3}. For standard WLI, we obtain, on average, $D=17.047\rm\, \frac{ps}{nm \cdot km}$ at $\lambda_0 \approx 1560.5\rm\,nm$ with a standard deviation of $\sigma_{\rm classical} = 0.051\rm\, \frac{ps}{nm \cdot km}$. This result is amongst the best reported to date in the literature~\cite{Hlubina_one_percent,Laurent_dispersion,Kardas_160fs,Hlubina_sub_100fs,Ye_few_percent,MethodsComparison,Galle_280fs}.
For Q-WLI, we measure, on average, $D=17.035\rm\, \frac{ps}{nm \cdot km}$ at $\lambda^*=1560.493\rm\,nm$ with a significantly better standard deviation of $\sigma_{N00N} = 0.021\rm\, \frac{ps}{nm \cdot km}$.

In our two sets of data, we observe a difference of $0.012\rm\, \frac{ps}{nm \cdot km}$ between the central values which is larger than the deviation expected from statistical uncertainties ($0.006\rm\, \frac{ps}{nm \cdot km}$).
Polarisation mode dispersion is also excluded as it would introduce at most an offset of $0.003\rm\, \frac{ps}{nm \cdot km}$.
Consequently, the difference in central values must originate from systematic errors.
In this sense we compute that, for standard WLI, the difference is explained by either a slight wavelength offset of the spectrometer ($<0.2\,\rm nm$), or by a slightly unbalanced interferometer ($\sim$1.5$\,\rm\mu m$). Both types of errors induce an error on the fitting parameter $\lambda_0$ which is translated to an error in $\frac{d^2n}{d \lambda^2}\Big|_{\lambda_0}$~\cite{Naganuma_subwavelength_1990,Diddams_dispersion-WLI_1996}.
At this point, we emphasise again that in our Q-WLI approach, $\lambda^*$ is essentially known with absolute accuracy and an unbalanced interferometer does not influence the measurement.
As Q-WLI presents less sources of systematic errors, it is therefore natural to consider that Q-WLI determines chromatic dispersion with absolute accuracy.

We further emphasise that our measurements performed with Q-WLI involve $\sim$62 times less photons transmitted through the SUT compared to standard WLI. It is therefore interesting to compare the achievable precision normalised to the number of transmitted photons. For each standard and quantum interferogram, the number of photons reaching the interferometer outputs is $N_{\rm std} \approx 2.0 \cdot 10^{10}$ and $N_{\rm quant} \approx 3.1 \cdot 10^8$, respectively. Consequently, the standard and quantum methods achieve precisions of $\left(\Delta D \right)_{{\rm std}} = 7146\,{\rm \frac{ps}{nm \cdot km}} / \sqrt{N_{\rm std}}$ and $\left(\Delta D \right)_{{\rm quant}} = 372\,{\rm \frac{ps}{nm \cdot km}} / \sqrt{N_{\rm quant}}$, respectively.
In other words, in addition to being more prone to systematic errors, the standard measurement requires $369 \times$ more photons for achieving the same precision as Q-WLI.

\subsection{Device calibration using Q-WLI}

Another advantage provided by Q-WLI lies in straightforward device calibration.
All the optical components in the interferometer actually show small residual chromatic dispersion, and this undesired offset needs to be evaluated and subtracted from the data in order to avoid systematic errors. In both cases, this implies performing a measurement without any SUT.\\
Note that in standard WLI, removing the SUT significantly unbalances the interferometer, and in order to observe interference the length of the reference arm has to be reduced accordingly (typically on the order of 1\,m). This procedure is technically challenging, time-consuming, and might lead to additional systematic errors.\\
At this point, Q-WLI shows its ability for user-friendly operation.
Even after removing the SUT, interference is observed, without any interferometer realignment.
\begin{figure}[h] 
\includegraphics{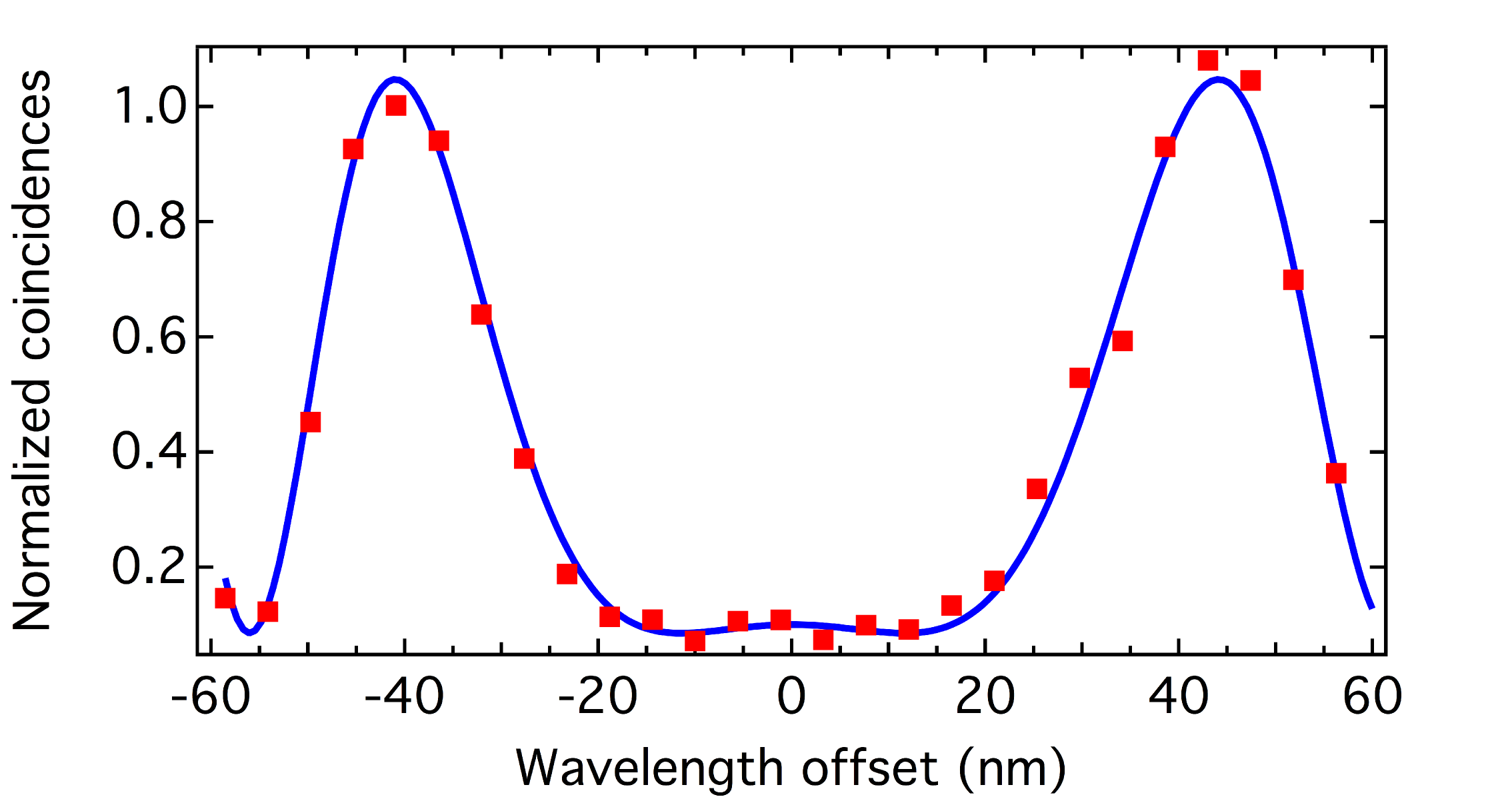}
\caption{Experimental results when using Q-WLI for inferring residual chromatic dispersion in our interferometer without the SUT. Red dots, data points; Blue lines, appropriate fit to the data.\label{Fig4}}
\end{figure}    
\figurename~\ref{Fig4} shows the experimental results that we have obtained when measuring chromatic dispersion in our bare interferometer, \textit{i.e.} without the SUT. It turns out that in our interferometer, residual chromatic dispersion amounts to $\sim$10\% of the measured values on the 1\,m SUT.
For all the data discussed above, except for the raw data in \figurename~\ref{Fig2}(a) and (b), we have subtracted the residual chromatic dispersion.

\section{Conclusions}

We have introduced and demonstrated the concept of spectrally-resolved Q-WLI exploiting energy-time entangled two-photon $N00N$-states.
Compared to standard measurements, the $N00N$-state permits achieving a two times higher phase sensitivity.
More strikingly, the peculiar use of such quantum states of light reduces the number of free parameters for fitting experimental data from three to one, representing a major advantage for determining optical properties with high precision and absolute accuracy.
In addition, our setup does not require a balanced interferometer for performing the measurement which represents a significant time-saving advantage compared to standard WLI. This is of particular interest for device calibration and when considering measurements on a large set of samples.

As an exemplary demonstration, we have applied our scheme to infer chromatic dispersion in a standard single-mode fibre, obtaining 2.4 times more precise results compared to state-of-the-art realisations, despite using $\sim$62 times less photons.

We note that the sensitivity of our approach could be further doubled by using a double-pass configuration~\cite{Laurent_dispersion}, towards achieving measurements on short samples, such as optical components and waveguide structures (mm to cm length scale).
Such measurements would also be of interest for medical applications where precise knowledge on chromatic dispersion in tissues is required to yield optimal image quality in optical coherence tomography~\cite{Drexler_OCT_1999}. In this perspective, the reduced number of photons required for quantum white-light interferometry is also highly interesting for measurements performed on photosensitive biological samples~\cite{Frigault_Live_2009,Celebrano_Single_Molecule_2011,Marek_Direct_2014}.
Regarding optical telecommunication systems, by rotating the polarisation of the entangled photon pairs, our setup could be also used for measuring polarisation mode dispersion in optical components, which would lead to refinement of manufacturing processes.

In addition, total measurement times could be reduced far below 1\,s by employing high-speed superconducting detectors showing $\sim$3 orders of magnitude higher saturation levels compared to the InGaAs SPDs employed here~\cite{Zwiller_detector_2017}. Alternatively, quantum-inspired strategies may also prove to be suitable~\cite{Mazurek_qinspired_2013,Manceau_qinspired_2017}.

In summary, we believe that combining fundamental and conceptual advantages enabled by quantum light is a very promising approach for the future development and improvement of applications requiring absolute and high-precision measurements of optical properties.

\section{Acknowledgements}

The authors acknowledge financial support from the Foundation Simone \& Cino Del Duca, the European Commission for the FP7-ITN PICQUE project (grant agreement No 608062), l'Agence Nationale de la Recherche (ANR) for the CONNEQT, SPOCQ and SITQOM projects (grants ANR-EMMA-002-01, ANR-14-CE32-0019, and  ANR-15-CE24-0005, respectively), and the iXCore Research Foundation.
We also thank M.~T. P\'erez Zaballos for help on statistical data treatment, as well as M. Mitchell, T. Debuisschert, A. Levenson, and P. Neumann for fruitful discussions.


\newpage
$\,$\\
\newpage

 \onecolumngrid
\begin{center}
{\Large Supplementary Information}\\
$\,$
\end{center}
 \twocolumngrid

\section{Mach-Zehnder interferometer stabilisation}

Without active interferometer phase stabilisation, we observe $2 \pi$ phase drifts every few seconds due temperature drifts in the laboratory. This limits severely the integration times for both the classical and quantum measurements. Therefore, we employ an active phase stabilisation system. It is made of an actively wavelength-stabilised 1560.5\,nm reference laser sent in the counter-propagating way through the interferometer, and a piezoelectric translation stage in the reference arm of the interferometer~\cite{Kaiser_source2}. 
The feedback loop has a bandwidth of 100\,Hz which results in a long-term phase stability of $<\frac{2\,\pi}{40}\,\rm rad$.

\section{Spatial and polarisation mode overlap}

In order to obtain high-visibility interference patterns at the interferometer output, the photon (or photon pair) contributions from both interferometer arms need to be made indistinguishable in both the spatial and polarisation modes. Spatial mode overlap is ensured by using a fibre-optic beam-splitter at the interferometer output and input~\cite{Lee_overlap}. Polarisation mode overlap is obtained using fibre-optic polarisation controllers in both interferometer arms.
These components are not shown in the main text figures in order to simplify the reading of the manuscript.

\section{Quality of the entangled photon pair source}

We infer the entanglement quality of our photon pair source in the following configuration. We fix the analysis wavelength of the spectrometer at 1550\,nm and post-select the desired $N00N$-state by a coincidence measurement. Then, the path length difference of the MZI is scanned and the two-photon coincidence rate is recorded. We measure sinusoidal oscillations with a raw fringe visibility of $87.1 \pm 2.2\%$ which increases to $95.5 \pm 2.6\%$ after the subtraction of detectors' dark counts. In other words, we obtain a fidelity of 97.8\% to the desired $N00N$-state. We explain imperfections by unbalanced losses between the two arms of the interferometer and multi-pair contributions.

\section{Normalisation of intensity and coincidence spectrograms}

For the classical strategy, normalisation is obtained by recording two reference spectrograms with either interferometer arm being blocked.
Normalisation is obtained by dividing the data by the sum of both reference spectrograms.

For the entanglement-enabled strategy, we normalise the coincidence counts by taking advantage of the \textit{undesired} contributions in which the paired photons take opposite paths inside the interferometer. These contributions do not interfere at the interferometer output, such that the related (non-zero time delay) coincidence rate is directly proportional to the spectral intensity of the photon pair generator.
Normalisation is obtained by dividing the $N00N$-state coincidences by two times the sum of the non-$N00N$-state coincidences.

\section{Home-made single-photon spectrometer}

As a single-photon spectrometer, we use a wavelength tunable motorised bandpass filter (Yenista XTM-50) followed by a low noise single-photon avalanche photodiode (id quantique id230) operated at 25\% detection efficiency. The transmission loss of the filter is measured to be 4\,dB such that the total quantum efficiency of the single-photon spectrometer is $\sim 10\%$.

\section{Derivation of the fitting function for standard WLI}

As outlined in the manuscript, the wavelength dependent phase shift at the interferometer output is
\begin{equation}
\phi(\lambda) = \frac{2\,\pi}{\lambda} \left( n(\lambda) \cdot L_{\rm s} - L_{\rm r} \right). \label{PhaseEquationOnePhoton}
\end{equation}
We approximate now $n(\lambda)$ by a third order Taylor series: $n(\lambda) = n(\lambda_0 + \Delta \lambda) \approx n(\lambda_0) + \sum_{k=1}^3 \frac{1}{k!}\, \frac{d^kn}{d \lambda^k}\Bigg|_{ \lambda_0} \cdot (\Delta \lambda)^k$.\\
This leads to
\begin{eqnarray}
\phi(\lambda_0 + \Delta \lambda) &\approx& 2\,\pi\,L_{\rm s} \Bigg( \frac{n(\lambda_0)}{\lambda_0 + \Delta \lambda} + \frac{d n}{d \lambda}\Bigg|_{ \lambda_0} \cdot \frac{\Delta \lambda}{\lambda_0 + \Delta \lambda} \nonumber\\
\,&\,&+ \frac{1}{2} \, \frac{d^2 n}{d \lambda^2}\Bigg|_{ \lambda_0} \cdot \frac{(\Delta \lambda) ^2}{\lambda_0 + \Delta \lambda}\\ \nonumber
\,&\,& + \frac{1}{6} \, \frac{d^3 n}{d \lambda^3}\Bigg|_{ \lambda_0} \cdot \frac{(\Delta \lambda) ^3}{\lambda_0 + \Delta \lambda} \Bigg)\nonumber\\
\,&\,& - \frac{2\,\pi\,L_{\rm r}}{\lambda_0 + \Delta \lambda}.
\end{eqnarray} 
Note that, in general, the interference fringes obtained at the interferometer output are usually too closely spaced to be resolved by a commercial spectrometer because  of the strong phase-dependence on zero and first order derivatives. In order to cancel these terms, the interferometer has to be precisely equilibrated to the so-called stationary phase point (SPP), which is found at $L_{\rm r} = \left( n(\lambda_0) - \frac{d n}{d \lambda}\Big|_{ \lambda_0} \cdot \lambda_0 \right) L_{\rm s}$.
Note that this point has to be found individually for each new sample with an accuracy on the order of a few micron.
After finding the SPP, the dominant term is $\frac{d^2n}{d \lambda^2}\Big|_{ \lambda_0}$ and the phase term simplifies to
\begin{eqnarray}
\phi(\lambda_0 + \Delta \lambda) &\approx& 2\,\pi\,L_{\rm s} \Bigg( \frac{1}{2} \, \frac{d^2 n}{d \lambda^2}\Bigg|_{ \lambda_0} \cdot \frac{(\Delta \lambda) ^2}{\lambda_0  + \Delta \lambda} \nonumber\\
&\,& + \frac{1}{6} \, \frac{d^3 n}{d \lambda^3}\Bigg|_{ \lambda_0} \cdot \frac{(\Delta \lambda) ^3}{\lambda_0 + \Delta \lambda} \Bigg) + \phi_{\rm off}, \label{ClassicalPhaseDerived}
\end{eqnarray}
in which $\phi_{\rm off} = 2\,\pi\,L_{\rm s} \, \frac{dn}{d \lambda}\Big|_{ \lambda_0}$ is a constant phase offset.\\
Assuming that $L_{\rm s}$ is known precisely, the required fitting parameters are therefore $\tfrac{d^2n}{d \lambda^2}\Big|_{ \lambda_0}$, $\tfrac{d^3n}{d \lambda^3}\Big|_{ \lambda_0}$ and $\lambda_0$.

\section{Data fitting up to $\tfrac{d^2n}{d \lambda^2}\Big|_{ \lambda_0}$ and $\tfrac{d^3n}{d \lambda^3}\Big|_{ \lambda_0}$}

Data obtained with standard WLI require a fitting function taking into account terms up to $\frac{d^3n}{d \lambda^3}\Big|_{ \lambda_0}$ to obtain the most precise and accurate results.

Fitting the data with the function described in equation~\ref{ClassicalPhaseDerived} leads to $D=17.047\rm\, \frac{ps}{nm \cdot km}$ at $\lambda_0 \approx 1560.5\rm\,nm$ and $\sigma_{\rm classical} = 0.051\rm\, \frac{ps}{nm \cdot km}$ after 100 measurements on the same standard single-mode fibre.

\begin{center}
\begin{figure}[h] 
\includegraphics{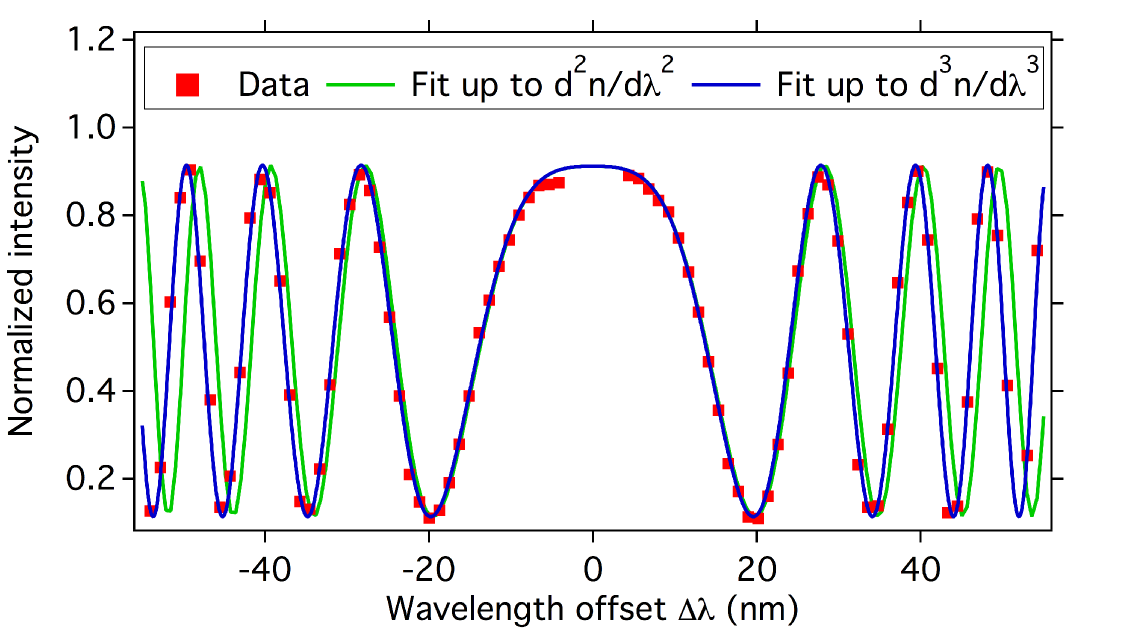}
\caption{Fitting standard WLI data (red dots) with a function taking into accound terms up to $\frac{d^2n}{d \lambda^2}\Big|_{ \lambda_0}$ (green line) and $\frac{d^3n}{d \lambda^3}\Big|_{ \lambda_0}$ (blue line). Only the later shows a perfect overlap with the experimental data.\label{Fig_supp}}
\end{figure}    
\end{center}

A fitting function taking into account only terms up to $\frac{d^2n}{d \lambda^2}\Big|_{ \lambda_0}$ is does not lead to a good overlap between data and experiment (see \figurename~\ref{Fig_supp}) which leads to both an offset and a larger standard deviation, \textit{i.e.} $D = 17.070 \pm 0.054\rm\,\frac{ps}{nm \cdot km}$.

\section{Derivation of the fitting function for Q-WLI}

For the Q-WLI, the phase term is given by the two-photon phase $\phi_{\rm N00N} = \phi(\lambda_1) + \phi(\lambda_2)$, which can be calculated using equation~\ref{PhaseEquationOnePhoton}.
Respecting the conservation of the energy, \textit{i.e.} $\frac{1}{\lambda_{\rm p}} = \frac{2}{\lambda^*} = \frac{1}{\lambda_1} + \frac{1}{\lambda_2}$, and setting $\lambda_2 = \lambda^* + \Delta \lambda$ leads to
\begin{eqnarray}
\phi_{N00N}(\lambda^* + \Delta \lambda) &\approx& 2\,\pi\,L_{\rm s} \cdot \Bigg( \frac{1}{2} \frac{d^2 n}{d \lambda^2}\Bigg|_{ \lambda^*} \cdot \frac{(\Delta \lambda)^2}{\frac{\lambda^*}{2}+\Delta \lambda} \nonumber \\
\,&\,&  + \frac{1}{6} \frac{d^3 n}{d \lambda^3}\Bigg|_{ \lambda^*} \cdot \frac{(\Delta \lambda)^4}{\left( \frac{\lambda^*}{2}+\Delta \lambda\right)^2}\Bigg) \nonumber \\
 \,&\,& + \phi_{\rm off}, \label{QuantumPhaseDerived}
\end{eqnarray}
where we consider the phase offset $\phi_{\rm off} = \frac{4\,\pi \left( n(\lambda^*) \,L_{\rm s} - L_{\rm r} \right) }{\lambda_0}$ to be constant thanks to the active phase stabilization system.

%

\end{document}